\documentclass{aastex631}
\shorttitle{CN 2-1 \&  CS 5-4 in  Arp 299}
\shortauthors{Junzhi Wang et al.}
\graphicspath{{./}{figures/}}

\begin{document}

\title{CN 2-1  and  CS 5-4  observations toward Arp 299 with the  SMA}

\author[0000-0001-6106-1171]{Junzhi Wang}
\affiliation{School of Physical Science and Technology, Guangxi University, Nanning 530004, People's Republic of China; junzhiwang@gxu.edu.cn}
\author{Chunhua Qi}
\affiliation{Harvard-Smithsonian Center for Astrophysics, 60 Garden Street, Cambridge, MA 02138}
\author{Shanghuo Li}
\affiliation{Korea Astronomy and Space Science Institute, 776 Daedeokdae-ro, Yuseong-gu, Daejeon 34055, Republic Of Korea}
\author{Jingwen Wu}
\affiliation{University of Chinese Academy of Sciences, Beijing 100049, People's Republic of  China}
\affiliation{National Astronomical Observatories, Chinese Academy of Sciences, Beijing 100101, People's Republic of China}

\begin{abstract}

Dense gas is the key for understanding star formation in galaxies. We present high resolution ($\sim3''$) observations  of   CN 2-1 and CS 5-4  as dense gas tracers toward Arp 299, a mid stage major merger of galaxies,  with  the  Submillimeter Array (SMA).  The spatial distribution of CN 2-1 and CS 5-4   are generally consistent with each other, as well as HCN 1-0 in literature.  However, different line ratios of   CS 5-4 and  CN 2-1 are found in A, B, and C regions, with highest value in B.
Dense gas fraction decreases from IC 694 (A), to NGC 3690 (B) and the overlap starburst region (C and C$'$), which indicates that  circum-nuclear  upcoming starburst in A and B will be more efficient than that in the overlap region of  Arp 299.

\end{abstract}


\section{Introduction} \label{sec:intro}

Galaxy-galaxy interactions and merging play an important role in the
 formation and  evolution of galaxies,  and can trigger
intense starburst activity and/or supply the fuel for AGN \citep{1972ApJ...178..623T,2005MNRAS.361..776S}.
Studying  molecular gas  in merging galaxies is important for
understanding  merging process,  triggering of nuclear and
off-nuclear starbursts, and the process of star formation in
extreme environments.  Imaged with good spatial resolution, nearby
major mergers allow detailed investigation of the exact location
of  molecular gas that directly fuels the star formation and/or
nuclear activity.

 Because of  low upper-level energy   and  low critical density ($\sim5\times10^2$cm$^{-3}$), low-$J$ (2-1, 1-0) transitions of CO
 lines  are not necessarily good  tracer of star-forming gas \citep{2004ApJ...606..271G,2012ApJ...745..190L},
 which is better traced by lines of  high-dipole moment molecules, such as HCN, HNC, HCO$^+$ and  CS \citep{2004ApJ...606..271G,2015ApJ...800...70S,2019ApJS..241...19I,2021MNRAS.503.4508L}, or high $J$ CO \citep{2003ApJ...588..771Y,2008ApJ...684..957S,2010ApJ...724.1336M}. HCN 1-0 had been used
 to measure dense gas in a sample of  galaxies to study the relation between star formation rate and dense gas mass \citep{2004ApJ...606..271G}.  High resolution observations of HCN lines had been done toward several local major mergers, including Arp 299 (IC694+NGC3690)  \citep{1997ApJ...475L.107A,1999A&A...346..663C, 2006PASJ...58..813I}, Arp 220 \citep{2015ApJ...800...70S}, NGC 6240 \citep{2015ApJ...800...70S}, VV 114 \citep{2018ApJ...863..129S},  NGC 3256  \citep{2018ApJ...855...49H}, and   NGC 4038 \& NGC 4039 (the Antennae galaxy) \citep{2019AJ....157..131B}.

However, it was questioned whether HCN was an ideal  tracer of
 dense gas in  star forming regions of LIRGs/ULIRGs as HCN can be excited
 by the mid-IR pumping  or shocks \citep{2007ApJ...656..792P}, which
 can enhance HCN emission. On the other hand, HCN can be transferred to
 HNC and CN in molecular clouds under certain conditions. For example, it
can be transferred to CN with the UV photons in
  PDRs, which can reduce the HCN flux despite the presence of dense molecular
  gas.  Since multiple line observations of HCN 1-0, HCO$^+$ 1-0, HNC 1-0, and CS 3-2 toward a sample of more than 100 galaxies showed  similar linear relationships between their luminosities  and infrared luminosities, with a factor of several in  variation of line ratios in different galaxies \citep{2021MNRAS.503.4508L}, it is hard and not that important to do correction of conversion factor from HCN  line luminosity to  dense gas mass  in individual  galaxy.  However,   if we would like  to probe the detailed dense gas distribution in galaxies,  it is essential  to observe other dense gas tracers, such as lines of CS, CN, HCO$^+$ and HNC,   to obtain
 a better and   unbiased view of the dense gas.

 High resolution observations of dense gas tracers other than HCN lines had been done toward  several nearby major mergers, including CS 7-6 in Arp 220 and NGC 6240 \citep{2015ApJ...800...70S},    HCO$^+$ 1-0, 3-2, and 4-3, HNC 1-0,  and CS 7-6  in VV 114 \citep{2018ApJ...863..129S},  HNC, HCO$^+$ and CS lines in NGC 3256 \citep{2018ApJ...855...49H}, and    HCO$^+$ 1-0 and HNC 1-0 in the Antennae galaxy \citep{2019AJ....157..131B}.  Due to different critical densities and possible different chemical abundances to that of HCN, the lines of other high dipole moment molecules showed different spatial distribution to that of HCN lines \citep{2015ApJ...800...70S,2018ApJ...863..129S,2018ApJ...855...49H,2019AJ....157..131B}.  Among those observed sources,   Arp 220,  NGC 6240 and  NGC 3256 are late-stage mergers,  while VV 114 is a mid-stage merger and the  Antennae galaxy is at early-stage.  Overlap starburst is  found  in  the  Antennae \citep{2004ApJS..154..193W,2019AJ....157..131B}, while nuclear starburst activities dominate star formation in these mid and  late-stage mergers in VV 114, Arp 220,  NGC 6240 and  NGC 3256.  Thus, Arp 299, as a mid-stage merger with both nuclear starburst and overlap region starburst \citep{2009ApJ...697..660A}, is an ideal source to study overlap starburst during merging process.

Arp299, at a distance of $\sim$42Mpc \citep{1999A&A...346..663C},  had been observed at high resolution with $^{12}$CO 1-0, $^{13}$CO 1-0 and  HCN 1-0 \citep{1997ApJ...475L.107A,1999A&A...346..663C,2006PASJ...58..813I}, HCO$^+$ 1-0 \citep{2006PASJ...58..813I}, $^{12}$CO 3-2 and 2-1  \citep{2012ApJ...753...46S}, and CN 1-0  \citep{2004ASPC..320....3A}.  HCN 1-0  showed not only different properties to that of CO 1-0 and 2-1, but also different to that of CN 1-0 \citep{2004ASPC..320....3A} and   HCO$^+$ 1-0 \citep{2006PASJ...58..813I}.  Based on the high resolution data,  most of  HCN 1-0 emission comes from the left
 galaxy IC694 \citep{1997ApJ...475L.107A,1999A&A...346..663C}, while   the strongest CN 1-0 emission, with similar critical density to that of HCN 1-0  \citep{ 2015PASP..127..299S},  is at the  overlap starburst region \citep{2004ASPC..320....3A}.  It is necessary to use other  dense gas tracers, such as CS lines,  to better probe  the dense molecular gas and determine the dense gas properties, including
 volume density, temperature, and chemistry in Arp 299.

In this paper,  we will present high resolution observational results of CN 2-1 and CS 5-4 toward Arp 299 with the  Submillimeter Array (SMA) \citep{2004ApJ...616L...1H}.
 Observations and Data reduction  will be described in \S2, results and discussion will be  present in \S3, and a brief summary will be made in \S4.

\section{Observations and Data reduction} \label{sec:obs}

The observations of  CN 2-1 and CS 5-4  toward Arp 299  were taken   with the SMA at compact configuration   (Project ID: 2007B-S015, PI: Junzhi Wang) on March 19 th and 20 th,   2008, respectively.  CN 2-1 line was set in the lower side band  with the local oscillator  (LO) frequency of 229.304 GHz, while CS 5-4 was in upper side band with LO frequency of 237.475 GHz.  Callisto was used as flux calibrator, while the  quasar J1924-292 was used as bandpass calibrator. Two  quasars, J1153+495 and  J0958+655, were used as  gain calibrators to correct  time dependent variations of both phase and amplitude.   Standard calibrations for  the uvdata  were made with  the IDL MIR package\footnote{https://github.com/qi-molecules/sma-mir}, while imaging and deconvolution with nature weighting  were made with CASA to obtain the final data cubes, after smoothing to  the velocity resolution of 20 km s$^{-1}$ for both  CN 2-1 and  CS 5-4.  The reference frequency  used for data cube of CN 2-1 is 224.415 GHz, which is calculated from CN 2-1 ($J$=$\frac{3}{2}-\frac{1}{2}$ $F$=$\frac{5}{2}-\frac{3}{2}$)   at rest frequency of 226.659543 GHz with $z$=0.01, and  242.510 
 GHz for CS 5-4 calculated from  rest frequency of  244935.556 GHz with  $z$=0.01.   The beam sizes are $3.12''\times2.98''$ with $PA=-12.25^\circ$  for  CN 2-1,  and $2.93''\times2.67''$ with $PA=-36.95^\circ$  for CS 5-4, while the typical rms noise levels are about 10 mJy beam$^{-1}$ for both   CN 2-1 and CS 5-4 at the velocity resolution of 20 km s$^{-1}$.  1.3 mm continuum distribution is also obtained with line free channels of both CN 2-1 and CS 5-4  data, which provide a map with beam size of $3.14''\times2.79''$, $PA=-33.37^\circ$  and rms level of 0.65 mJy beam$^{-1}$.

\section{Results and discussion} \label{sec:results}

As shown in  Figure 1(a) and (b), 1.3 mm continuum emission in Arp 299 can be found in IC 694 (A), NGC 3690 (B) and overlap starburst region (C and C$'$). The coordinates of  symbols `+', which indicate the locations of A, B, C and C$'$,   in Figure 1 and 2, are from 8.4 GHz continuum  \citep{2022A&A...658A...4R}.   The velocity integrated CN 2-1 emission is mainly from IC 694 (A) with peak emission at about 20$\sigma$ level, while weak emission can also be seen near NGC 3690 (B) and the overlap starburst region (C and C$'$) at 3 to 4$\sigma$ level.  CS 5-4  emission is also mainly from IC 694 (A)  with peak emission at about 8$\sigma$ level, while about 3$\sigma$ signal is also detected near B region.  The velocity integrated CN 2-1 and CS 5-4 overlaid on 1.3 mm continuum are presented in Figure 1(a) and (b), while spectra of CN 2-1  and CS 5-4 in A, B, and C regions  are presented in Figure 1(c-e).  A narrow CN 2-1 emission feature around C$'$ region is found at the velocity around -100 km s$^{-1}$. The velocity integrated map of this velocity component and the spectra of CN 2-1 and CS 5-4, are presented in Figure 2(a) and (b), respectively.

 The spatial distributions  of both CN 2-1 and CS 5-4   are generally consistent  with  that of HCN 1-0 \citep{1997ApJ...475L.107A,1999A&A...346..663C,2006PASJ...58..813I} with strong emission in A and weak emissions in B and C(C$'$) regions, while  the strong CN 1-0 emission in the overlap starburst regions of Arp 299  presented in \cite{2004ASPC..320....3A} seems not to be real signal.   On the other hand,   A and B+C regions had similar total CO 1-0 flux  \citep{1992ApJ...387L..55S}, which were confirmed with high resolution observations \citep{1997ApJ...475L.107A,1999A&A...346..663C}.  CO 2-1 showed similar spatial distribution to that of CO 1-0 \citep{2012ApJ...753...46S}.   Single dish observations of HCO$^+$ and HCN 1-0 lines with IRAM 30meter telescope toward A and B+C regions  also showed stronger emission in A  than that of B+C regions with   flux ratios between A and B+C regions of about 1.55 for HCO$^+$ 1-0 and 1.76 for  HCN 1-0  \citep{2011MNRAS.418.1753J}.  

   Amount of molecular gas in the overlap starburst regions (C and C$'$) was further supported by high resolution observations of optically thin $^{13}$CO 1-0 \citep{1997ApJ...475L.107A,1999A&A...346..663C} and $^{13}$CO 2-1 \citep{2012ApJ...753...46S}. The three main components (A, B, C+C$'$)  in Arp 299  have different gas properties.  There are strong $^{12}$CO emissions and amount of molecular gas traced by $^{13}$CO 1-0 and 2-1, most of which are with high volume density traced by dense gas tracers in  IC 694 (A), while  only strong $^{12}$CO,  $^{13}$CO and HCO$^+$ 1-0 emission without strong emission of tracers with critical density higher than that of HCN 1-0 were detected  in the overlap starburst regions (C and C$'$), which means that dense gas fraction is much less in  C and C$'$ than  A.  On the other hand, even though strong $^{12}$CO 1-0  \citep{1997ApJ...475L.107A,1999A&A...346..663C}  and 2-1  \citep{1999A&A...346..663C, 2012ApJ...753...46S} were detected in NGC 3690 (B), the molecular gas is much less than that of IC 694 and the overlap starburst regions, based on $^{13}$CO 1-0 \citep{1997ApJ...475L.107A,1999A&A...346..663C} and $^{13}$CO 2-1 \citep{2012ApJ...753...46S} observations.  
   
  There are several hyperfine  lines of CN 2-1 mainly in two groups, near 226.659543 GHz and 226.874764 GHz \citep{1985ApJS...58..341S}. Thus, the line widths of CN 2-1 are broader than that of CS 5-4 in Arp 299 (see Figure 1), because of the confusion of  CN 2-1 hyperfine lines.   CN 2-1 is stronger than CS 5-4 in A and C regions (see Figure 1), while they are comparable in B region, with low signal to noise ratio (see Figure 1 and Table 1).  A narrow CN 2-1 emission feature above 5 $\sigma$ with line width (FWHM) of less than 100 km s$^{-1}$ was found in C$'$ region, while no CS 5-4 emission was detected there (see Figure 2).  With comparable CO emission in C and C$'$ regions  \citep{1997ApJ...475L.107A,1999A&A...346..663C}, the less CS 5-4 and CN 2-1 in C$'$ than C region  indicates that the dense gas fraction in C$'$  is the lowest among the main gas components of A, B, C, and C$'$ regions,  in Arp 299. 
 

   The lack of HCN 1-0 \citep{1997ApJ...475L.107A, 2006PASJ...58..813I}, CN 2-1 and CS 5-4 emission at B, C and C$'$ regions with interferometer observations,   indicates that there are weak extended emission of these dense gas tracers, because comparable emissions of dense gas tracers exist in B and C regions to that of  region A based on the  observational results with  single dish telescope \citep{2011MNRAS.418.1753J}. The HCN 1-0 flux in  B, C and C$'$ regions picked up by interferometer observation  \citep{2006PASJ...58..813I} was only about 4 Jy km s$^{-1}$, which was about 50\% of that detected with single dish telescope \citep{2011MNRAS.418.1753J}. With more sensitive interferometer observation \citep{1999A&A...346..663C}, almost all the HCN 1-0 flux was picked up.    Further sensitive observations  with millimeter interferometers or large single dish telescopes  for lines with critical density higher than that of HCN 1-0, such as CN 2-1 and CS 5-4, are necessary to derive spatially resolved distribution of weak extended dense gas in Arp 299.  

Dense gas fraction in different regions had been determined by line ratios of  CO/HCN 1-0 \citep{1999A&A...346..663C}, which were 7$\pm1$ in A, 20$\pm1$ in B, 45$\pm8$ in C, and 29$\pm2$ in C$'$ regions.  So, if the  critical density  of HCN 1-0 is used as the threshold of dense gas, the dense gas fraction is the highest in A and lowest in C(C$'$) , while in between in B regions. 
With several times lower critical density than that of HCN 1-0, HCO$^+$ 1-0 showed different properties than that of  HCN 1-0 \citep{2006PASJ...58..813I}, with  HCO$^+$/HCN 1-0 ratio of 1.2 in A, 4.0 in B, and 2.1 in C regions.
  However, if the  critical density  of HCO$^+$ 1-0 is used as the threshold, A and B regions do have similar dense gas fraction  with CO/HCO$^+$ 1-0 ratio of $\sim5.8$ in A and $\sim$5.0 in B, while C(C$'$)  region  with  CO/HCO$^+$ 1-0 ratio of $\sim 13.8$  to $\sim $21.4 has  lower  dense gas fraction than that of A and B regions.

The critical density of CN 2-1 is several times higher than that of HCN 1-0 and several times lower than that of CS 5-4  \citep{ 2015PASP..127..299S}. Even though the  CN 2-1 and CS 5-4 data are not sensitive enough to present high signal to noise ratio detections in B and C regions with velocity integrated map (see Figure 1), spectra in individual regions can still provide useful information to obtain line ratios of CS 5-4/CN 2-1 (see also Figure 1), just similar to what had been done for HCN/HCO$^+$ 1-0 in  \cite{2006PASJ...58..813I}. The velocity integrated intensities of CN 2-1 and CS 5-4   in A, B, and C regions are listed in Table 1. CS 5-4 was not detected in C region. Even though dense gas fraction in B region derived by HCN 1-0  to CO 1-0 ratio is about 1/3 of that in A region  \citep{1999A&A...346..663C},  the line ratio between CS 5-4 to CN 2-1 in B region is about 3 times  of that in  A region, which means the fraction of  molecular gas  with density higher than the critical density of CS 5-4 in B region is comparable to that in A region. On the other hand, with CN 2-1 emission and non-detection of CS 5-4 in C(C$'$) regions, the fraction of molecular gas with density higher than the critical density of CS 5-4 in C(C$'$) regions is lower than that of A and B regions.  Dense gas fraction with different threshold traced by different dense gas tracers, including, HCO$^+$ 1-0, HCN 1-0, CN 2-1, and CS 5-4, provided slightly different properties in A, B, and C(C$'$) regions. We suggest that the differences are mainly caused by volume density instead of astrochemical abundances of different tracers.

With generally consistent  distribution of different dense gas tracers (HCN 1-0, CN 2-1, and CS 5-4) in Arp 299, the overlap starburst region (C and C$'$)  do have lower dense  gas fraction  than that of A and B regions supporting that starburst is stronger near the two nuclei than that of other regions.  Similar results of higher dense gas fraction near two nuclei than other regions were also found for one early stage merger, the  Antennae galaxy \citep{2019AJ....157..131B}. The efficient gas density increment near the nucleus due to the loss of angular momentum during merging process  \citep{1972ApJ...178..623T} is more important than the overlap starburst activity, even though active overlap starburst can be found in some early and mid stage mergers.  The overlap starburst may not be a common phenomenon during merging process of two gas-rich spirals, since there are many mergers without overlap starburst. For Arp 299, because the asymmetric molecular gas distribution in IC 694  revealed by CO lines \citep{1997ApJ...475L.107A,1999A&A...346..663C}, with the lack of molecular gas in the northeastern part of IC 694, the most possible way to form the overlap  C and C$'$ starburst regions was  that NGC 3690 went across the  northeastern part of IC694 during the merging process, which brings the gas from  NGC 3690 and the northeastern part of IC 694 to    C and C$'$ starburst regions. Such process can explain the lack of molecular gas in   the northeastern part of IC 694, much less molecular gas in NGC 3690 than that in IC 694, as well as  the location of overlap starburst region, which  is not in between the two galaxies (IC 694 and NGC 3690).  The star formation activity traced by  mid-IR emission at 38${\mu}$m also presented similar properties to that of dense gas tracers in A, B, and C(C$'$) regions, with 46\% of the total flux density in A region, 22\% in B region, 8\% in C(C$'$) in C region, and the remaining 24\% as diffuse emission   \citep{2002ApJ...571..282C}. Thus, both the upcoming star formation traced by dense gas tracers and on-going star formation traced by infrared emission  indicate that the  activities  of star formation rank from high to low as A, B, and C(C$'$) regions.

\begin{figure*}
\gridline{\fig{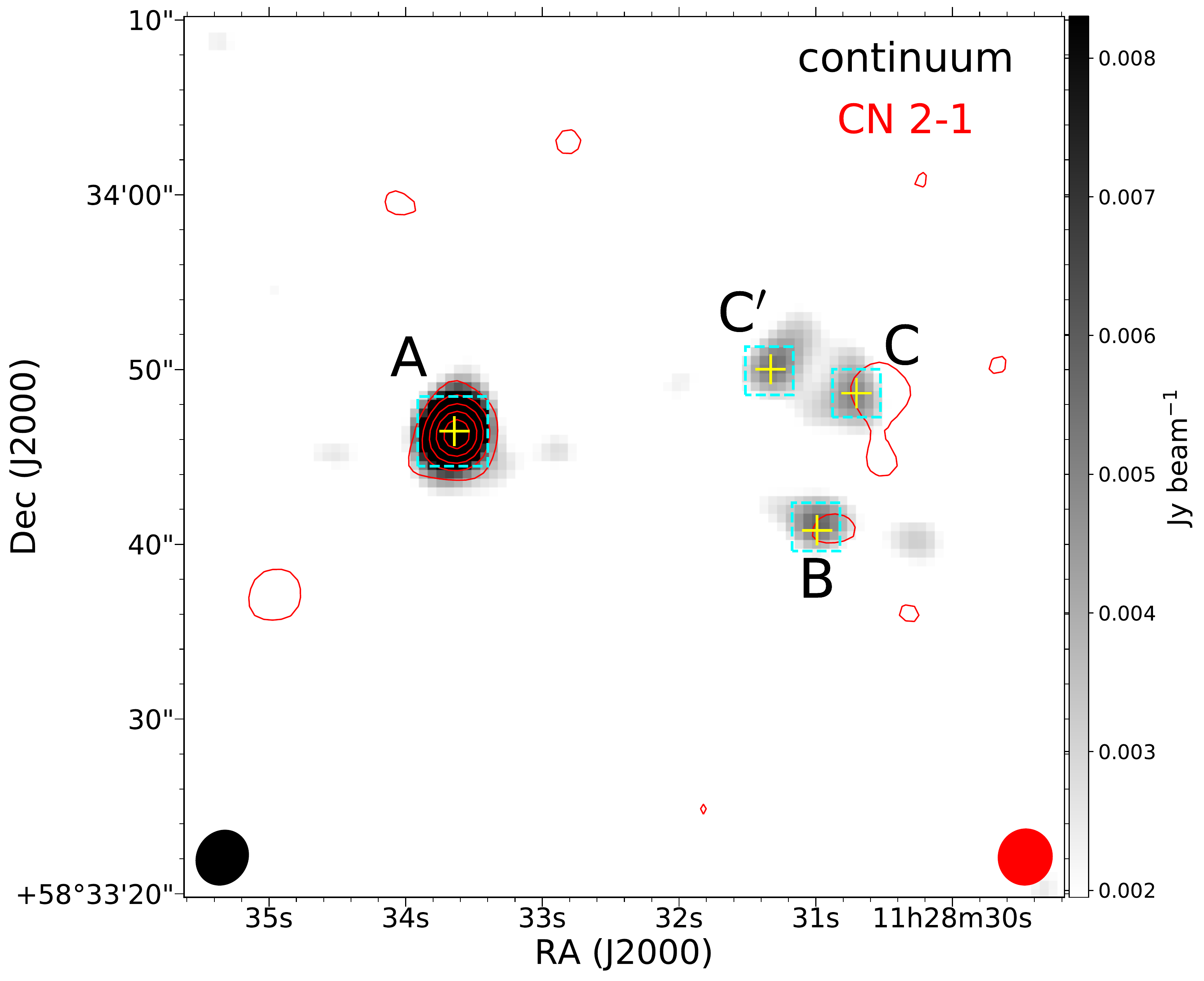}{0.5\textwidth}{(a)}
          \fig{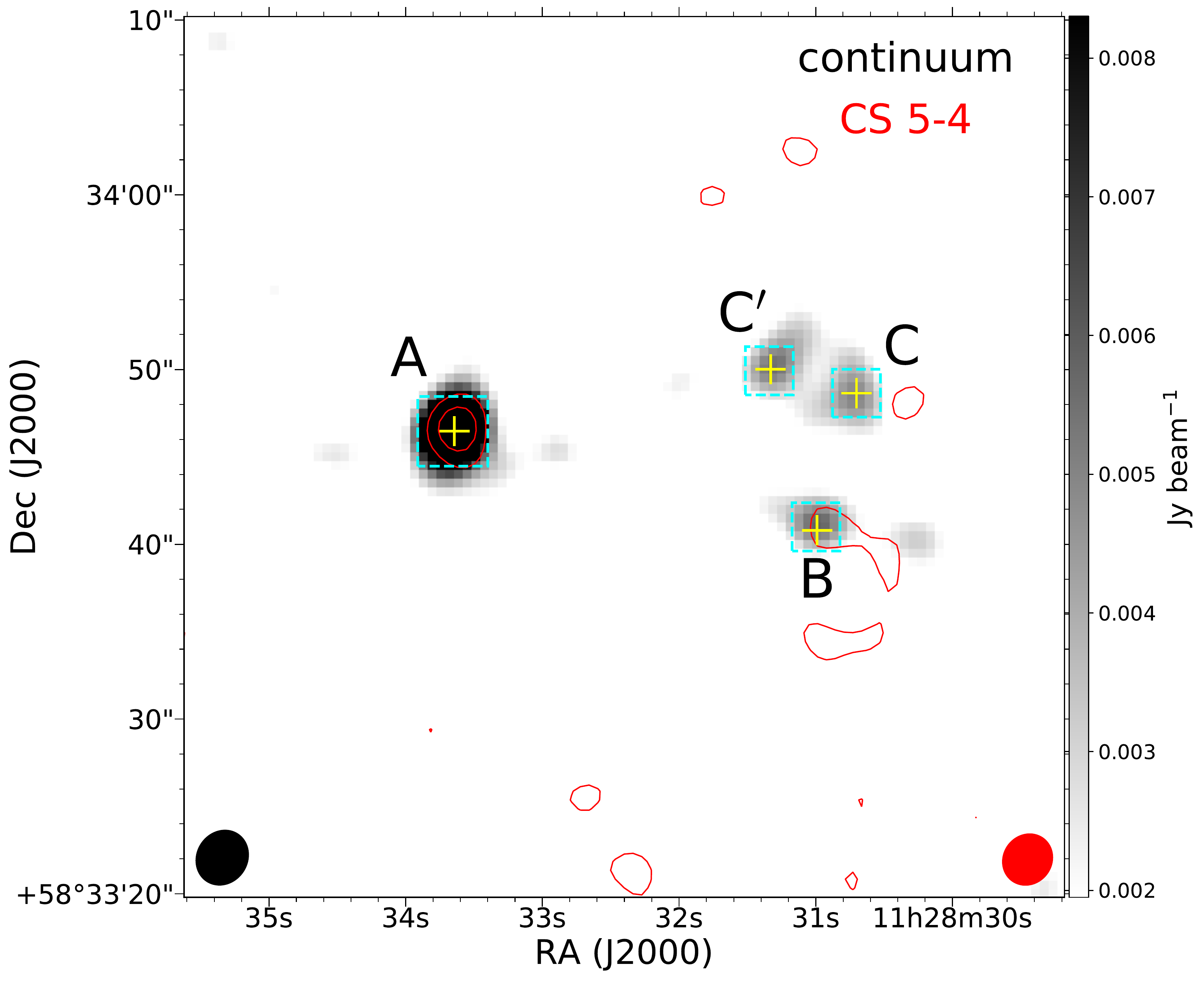}{0.5\textwidth}{(b)}
              }
\gridline{\fig{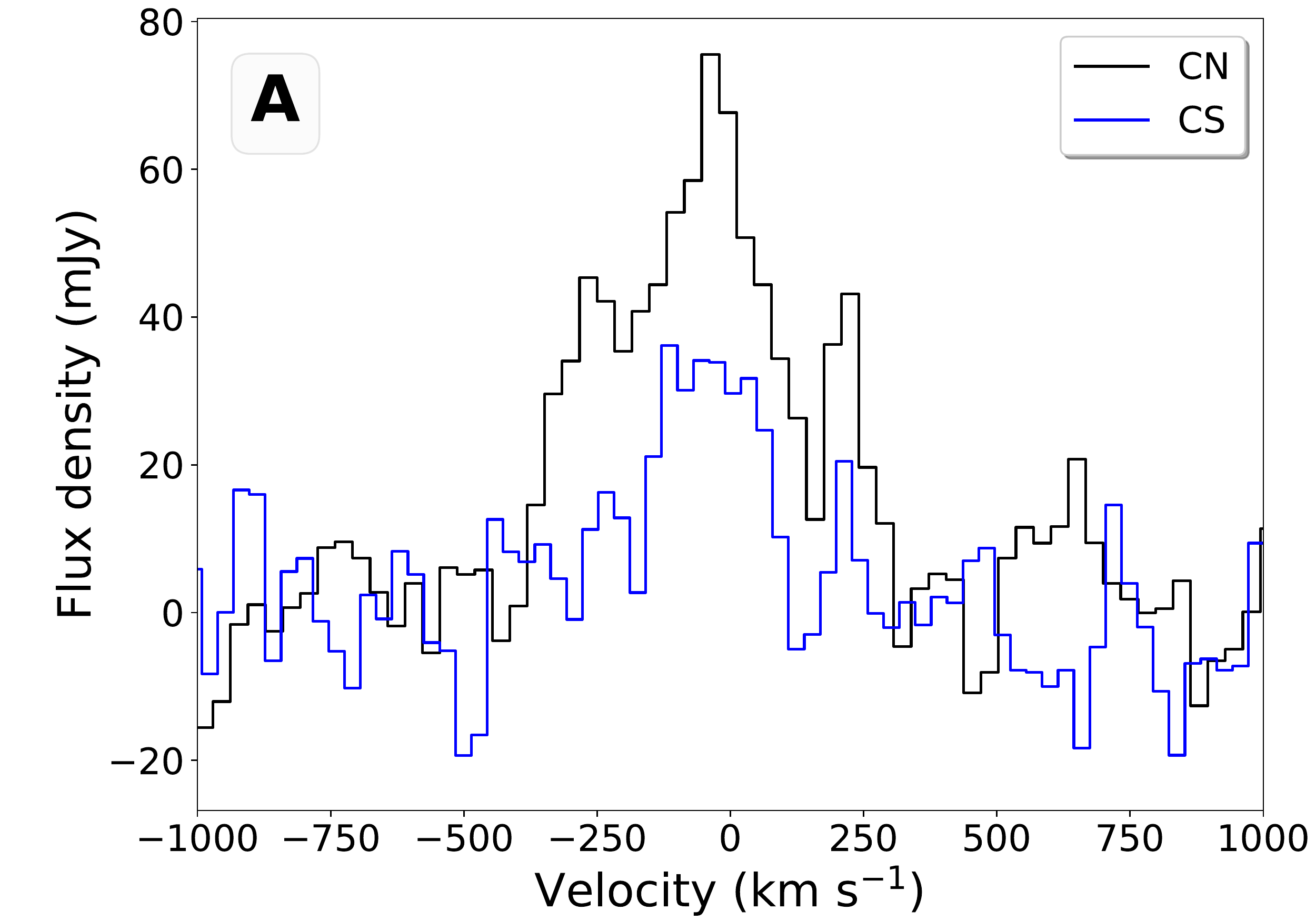}{0.33\textwidth}{(c)}   
\fig{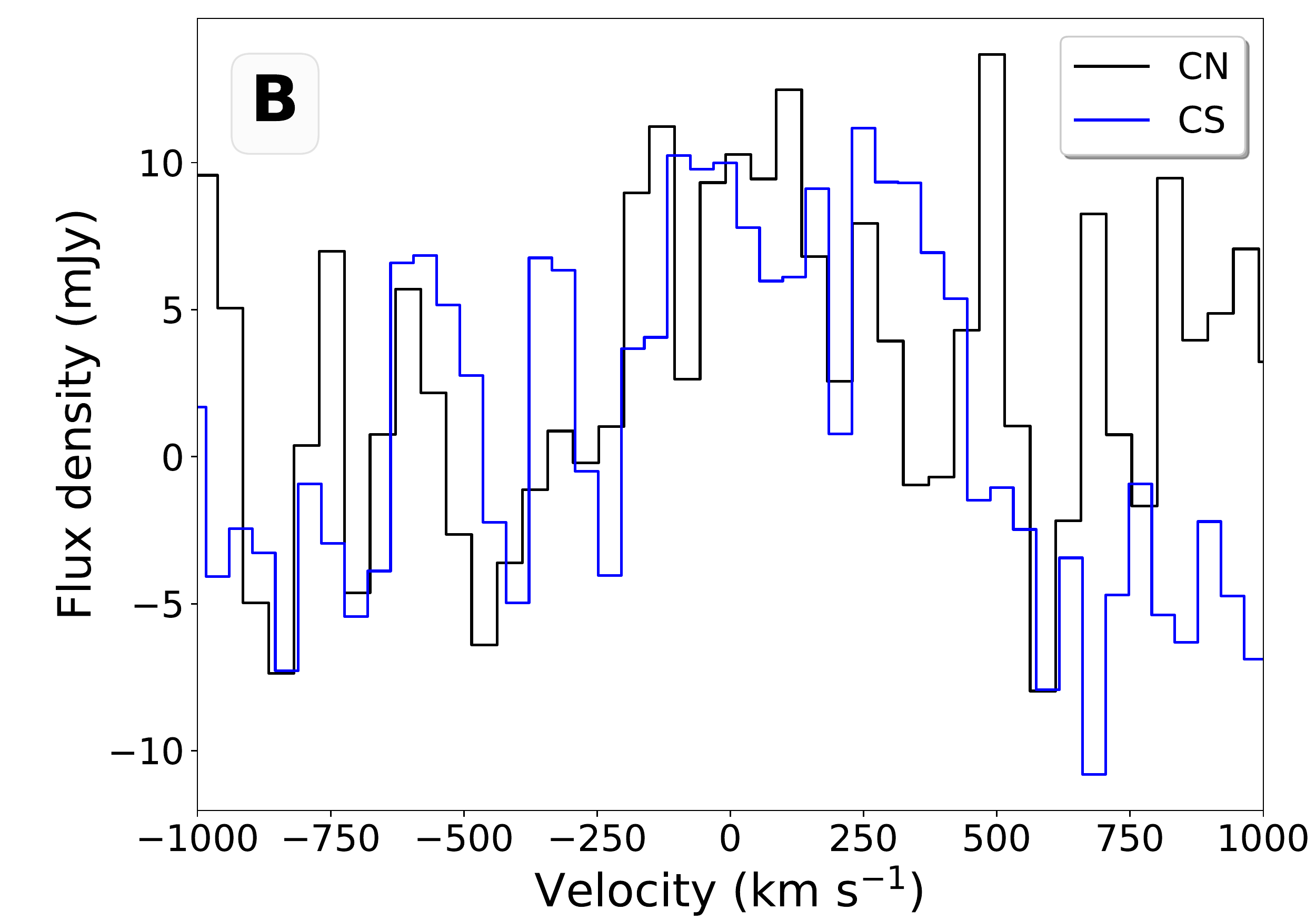}{0.33\textwidth}{(d)}
          \fig{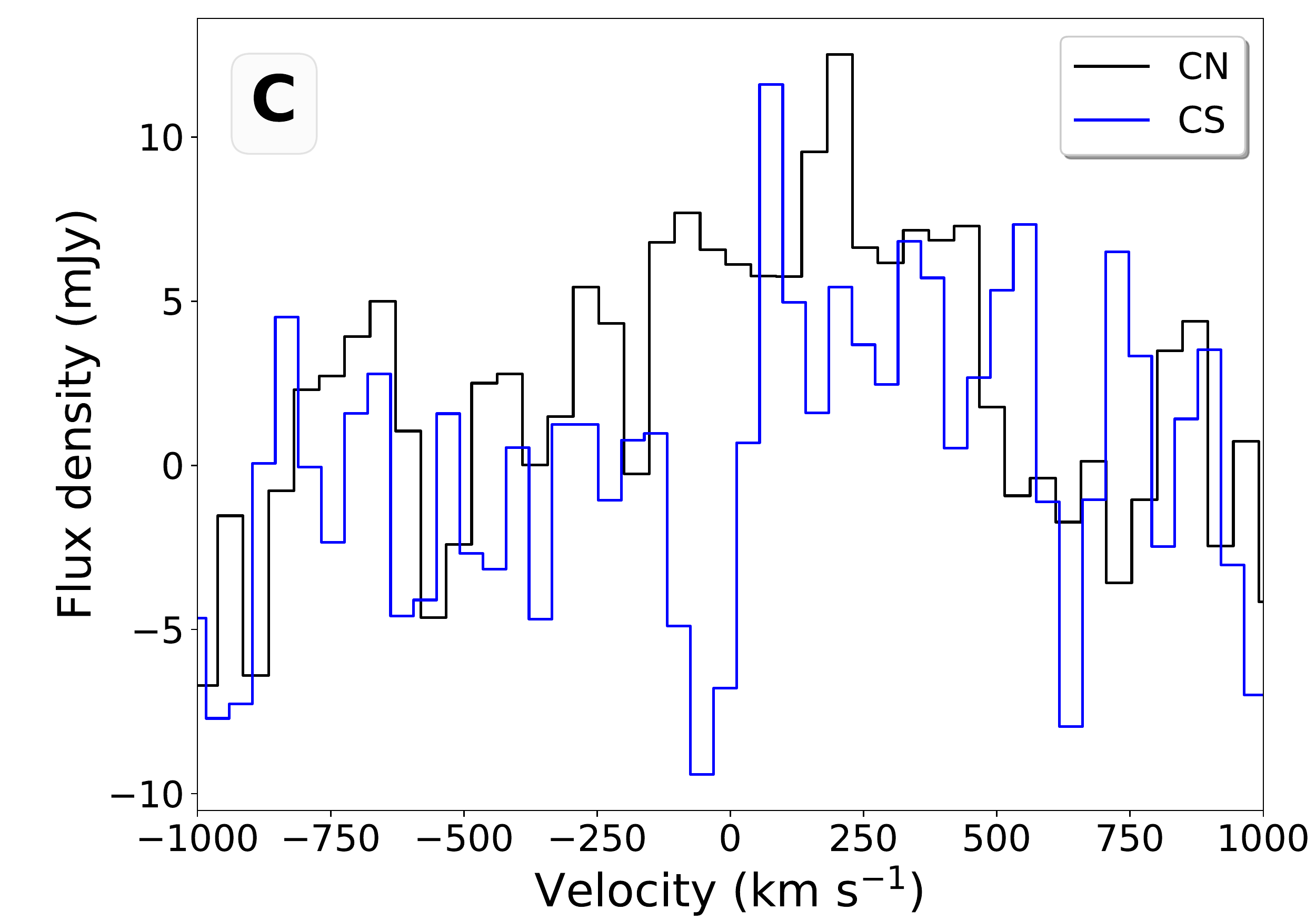}{0.33\textwidth}{(e)}
          }
\caption{a. Spatial distribution of velocity integrated  CN 2-1 (red contour) from -380 to 500 km s$^{-1}$  overlaid on 1.3mm continuum (greyscale).  The contour levels are 4, 8, 12, 16, 20, 24$\sigma$, while  1$\sigma$ is 1.33 Jy km s$^{-1}$ beam$^{-1}$ and the beam is 3.12$''\times2.98''$ with  $P.A.$=-15.25$^{\circ}$ shown at the bottom right corner.  The beam of 1.3mm continuum  shown at the bottom left corner  is 3.14$''\times2.79''$ with $P.A.$= -33.37$^{\circ}$, while 1$\sigma$ is 6.5$\times10^{-4}$Jy. b.  Velocity integrated CS 5-4 (red contour) from -180 to 500 km s$^{-1}$  overlaid on 1.3mm continuum (greyscale). The contour levels are  3, 6, 9$\sigma$, while  1$\sigma$ is 1.22 Jy km s$^{-1}$  beam$^{-1}$ and the beam is 2.93$''\times2.67''$ with  $P.A.$=-36.95$^{\circ}$ shown at the bottom right corner. c. Spectra of CS 5-4 and CN 2-1 within dashed box of  region A. d.  Spectra of CS 5-4 and CN 2-1 within dashed box of  region B. e. Spectra of CS 5-4 and CN 2-1 within dashed box of  region C. 
\label{fig:fig1}}
\end{figure*}

\begin{figure*}
\gridline{\fig{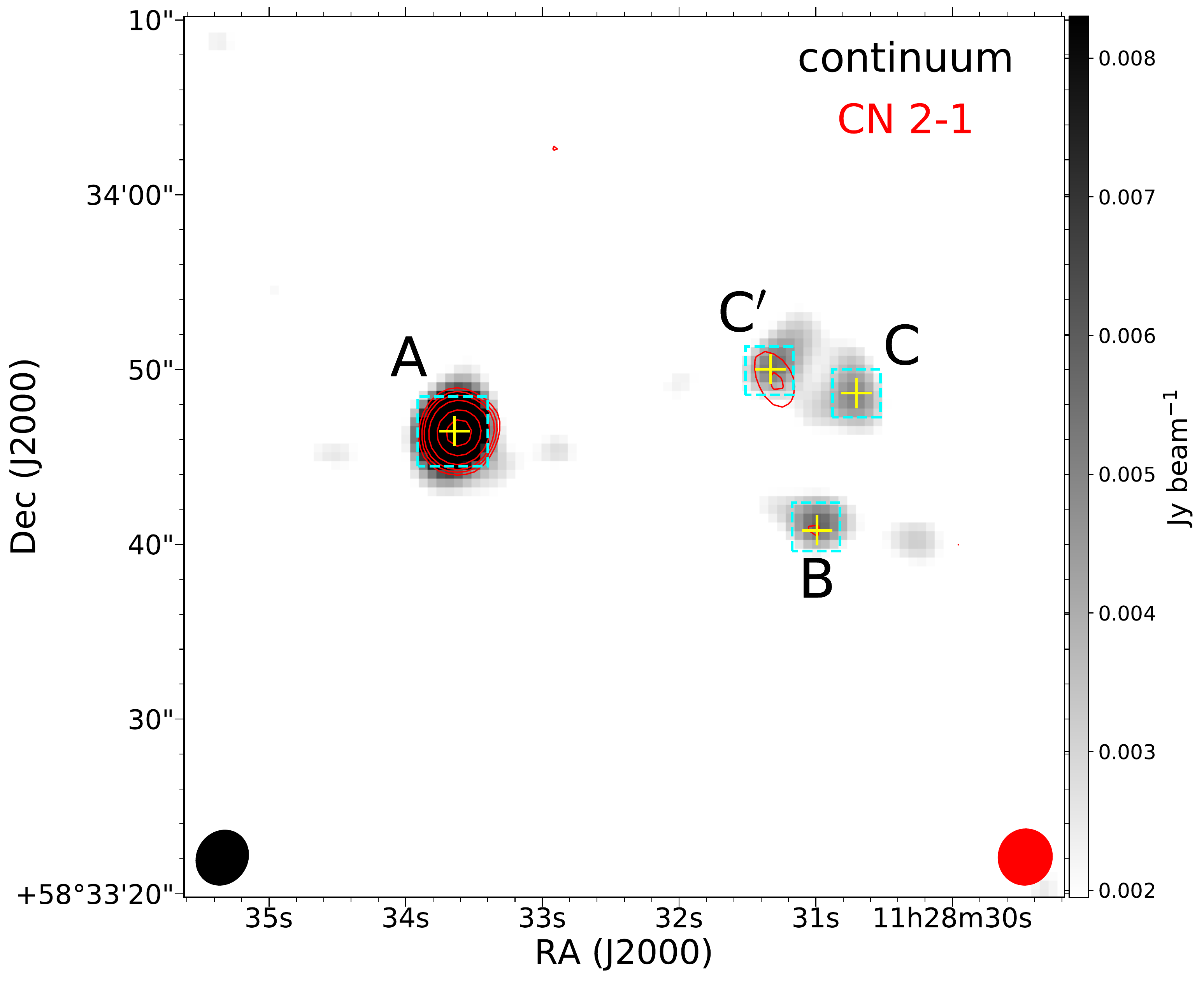}{0.5\textwidth}{(a)}
          \fig{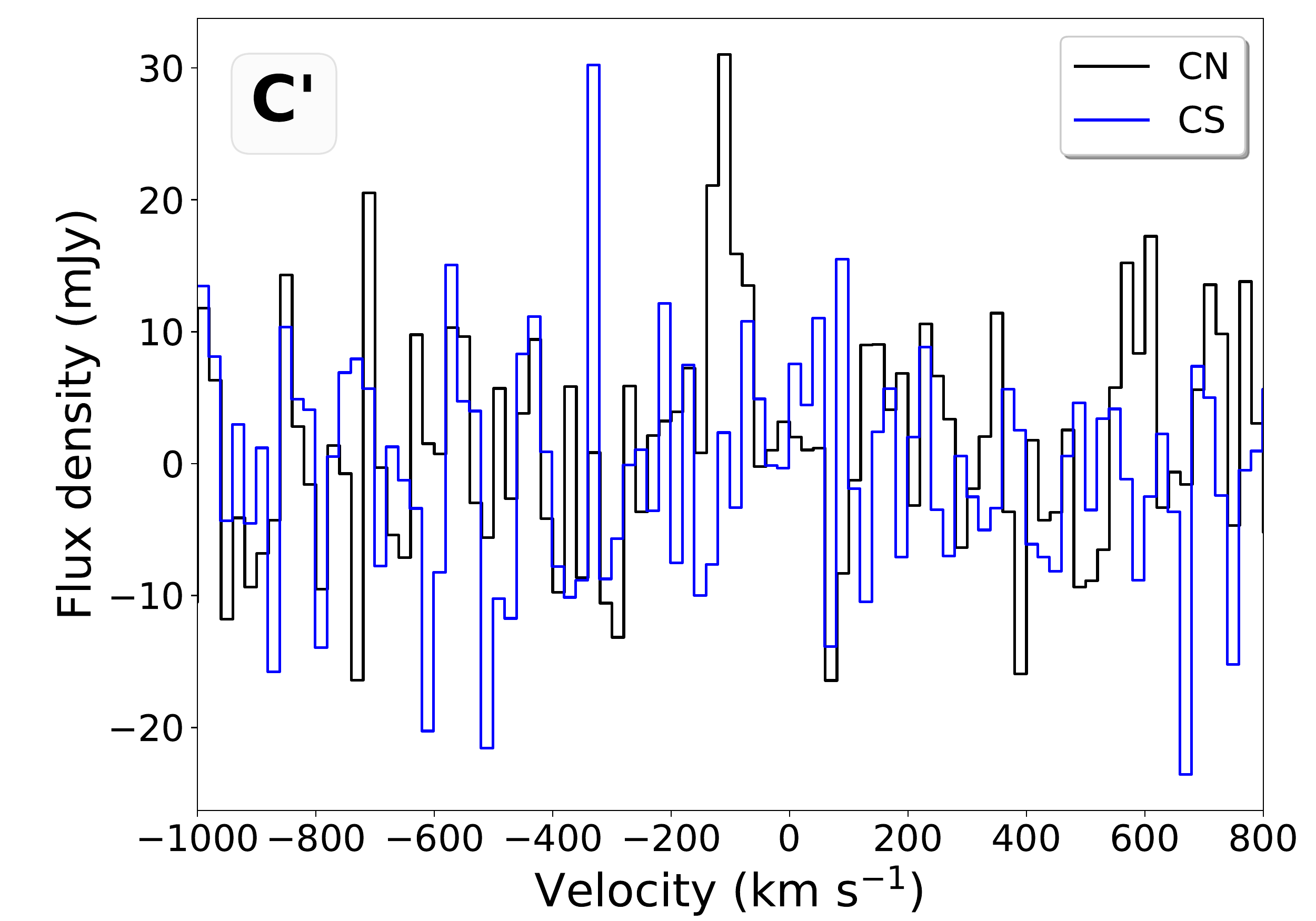}{0.5\textwidth}{(b)}
          }
\caption{a.Velocity integrated  CN 2-1 (red contour) from -180 to -40 km s$^{-1}$  overlaid on 1.3mm continuum (greyscale). The contour levels are 4, 5, 6, 8, 12, 16, 24  $\sigma$, while  1$\sigma$ is 0.53 Jy km s$^{-1}$ beam$^{-1}$, with the same beam parameters described in Figure 1. b.  Spectra of CS 5-4 and CN 2-1 within dashed box of  region C$'$.
\label{fig:fig2}}
\end{figure*}

\begin{table*}[b]
\begin{center}{Table 1. CN 2-1 and CS 5-4 velocity integrated intensities in A, B, and C regions. }\end{center}
\small
\begin{center}
\begin{tabular}{cccccc}
\hline
 Lines & A    & B &C  \\
\hline                                                            
CN 2-1  & 26.9$\pm$1.0  &3.9$\pm$1.2 & 5.0$\pm$1.4\\
CS 5-4 & 9.3$\pm$0.9  &4.3$\pm$1.2  & $<$3.3 (3$\sigma$)\\
\hline
\end{tabular}
\end{center}
Note:  The velocity integrated intensities are based on the spectra shown in Figure 1, with the units of Jy  km s$^{-1}$. 

\end{table*}

\section{Summary} \label{sec:sum}

With high resolution observations of CN 2-1 and  CS 5-4 toward a mid-stage major merger Arp 299 with the SMA, we obtained spatial distribution of dense gas at about 3$''$ level.  Most of CN 2-1 and CS 5-4 emissions are detected in the nuclear region of IC 694 (A), while only marginal detection of CN 2-1 and CS 5-4 are seen in NGC 3690 (B)  and the overlap starburst region (C and C$'$).  Highest CS 5-4 to CN 2-1 line ratio was found in B region among A, B and C regions of Arp 299. The dense gas fraction in different regions can vary with different threshold, observed with tracers with different critical densities, such as HCO$^+$ 1-0, HCN 1-0, CN 2-1, and CS 5-4. Dense gas fraction is the lowest in C(C$'$) region, when dense gas tracers with  critical densities higher than that of HCN 1-0, are used.
 The dense gas fraction decreases from A to B, C, and C$'$ regions in Arp 299. The lack of CN 2-1 and CS 5-4 emissions in B and C+C$'$ regions, together with strong emission of dense gas tracers there  detected with single dish telescopes,  implies there are   weak extended emission of dense gas tracers in B and C+C$'$ regions, which is similar to that of HCN 1-0 seen in \cite{1999A&A...346..663C}.

\begin{acknowledgments}
We thank Prof. Yu Gao for helpful discussion. This work is supported by  National Key Basic Research and Development Program of China (grant No. 2017YFA0402704) and the National Natural Science Foundation of China grant 12173067. We thank the anonymous referee for helpful comments, which improved the manuscript.  The Submillimeter Array is a joint project between the Smithsonian Astrophysical Observatory and the Academia Sinica Institute of Astronomy and Astrophysics and is funded by the Smithsonian Institution and the Academia Sinica.

 \end{acknowledgments}

\vspace{5mm}
\facilities{SMA}

\software{MIR, CASA}





\bibliography{Arp299_CS}{}
\bibliographystyle{aasjournal}



\end{document}